\documentclass[conference]{IEEEtran}

\usepackage{amsmath}
\usepackage{amssymb}

\usepackage{color}
\usepackage{dcolumn}
\usepackage{float}
\usepackage{graphicx}
\usepackage[latin9]{inputenc}
\usepackage{rotating}
\usepackage{psfrag}
\usepackage{tabularx}
\usepackage[hyphens]{url}
\usepackage{fancyvrb}

\usepackage[bookmarks, bookmarksopen, bookmarksnumbered]{hyperref}
\usepackage[all]{hypcap}
\newcommand{\ccl}{Cactus Configuration Language}

\begin{document}

\title{Component Specification in the Cactus Framework: The Cactus Configuration Language}

\author{
\IEEEauthorblockN{Gabrielle Allen}
\IEEEauthorblockA{Center for Computation \& Technology\\
Department of Computer Science\\
Louisiana State University\\
Baton Rouge, Louisiana 70803\\
Email: gallen@cct.lsu.edu}
\and
\IEEEauthorblockN{Tom Goodale}
\IEEEauthorblockA{7 Constantine Avenue, Heswall\\
Wirral CH60 5SU\\
United Kingdom\\
Email: goodale@gmail.com}
\and
\IEEEauthorblockN{Frank L\"offler}
\IEEEauthorblockA{Center for Computation \& Technology\\
Louisiana State University\\
Baton Rouge, Louisiana 70803\\
Email: knarf@cct.lsu.edu}
\and
\IEEEauthorblockN{David Rideout}
\IEEEauthorblockA{Perimeter Institute for Theoretical Physics\\
31 Caroline St.\ N.\\
Waterloo, Ontario N2L 2Y5\\
Canada\\
Email: drideout@perimeterinstitute.ca}
\and
\IEEEauthorblockN{Erik  Schnetter}
\IEEEauthorblockA{Center for Computation \& Technology\\
Department of Physics \& Astronomy\\
Louisiana State University\\
Baton Rouge, Louisiana 70803\\
Email: schnetter@cct.lsu.edu}
\and
\IEEEauthorblockN{Eric L. Seidel}
\IEEEauthorblockA{City College of New York\\
New York, New York 10031\\
Center for Computation \& Technology\\
Louisiana State University\\
Baton Rouge, Louisiana 70803\\
Email: eseidel01@ccny.cuny.edu}
}

\maketitle

\begin{abstract}
%\boldmath
%Complex frameworks are increasingly important to support the computational demands of 
%modern science. It is essential however to hide the complexity as much as possible, and present
%users with a simple, yet powerful, interface to the framework. This allows new users to quickly
%adapt their own code to be used with the framework and focus more on the science.
Component frameworks are complex systems that rely on many layers of abstraction to function
properly. One essential requirement is a consistent means of describing each individual component
and how it relates to both other components and the whole framework. As component frameworks
are designed to be flexible by nature, the description method should be simultaneously powerful, 
lead to efficient code,  
and be easy to use,
so that new users can quickly adapt their own code to work with the
framework.  

In this paper, we discuss the Cactus Configuration Language (CCL) which is used to describe 
components (``thorns'') in the Cactus Framework. The CCL provides a description language
for the variables, parameters, functions, scheduling and compilation of a component and 
includes concepts such as {\it interface} and {\it implementation} which allow thorns providing the same 
capabilities to be easily interchanged. We include several application examples 
which illustrate how community toolkits use the CCL and Cactus and identify
needed additions to the language.
\end{abstract}


% \fi
\IEEEpeerreviewmaketitle

% FL: I think we all agree that Erik deserves to be an author, so I comment
% this now.
%\todo{ES: I didn't help much with writing this paper, I didn't have
%  the time in the past days and won't have much time in the coming
%  days.  I don't think I should be an author.}

\section{Introduction}

Component frameworks provide a mechanism for efficiently developing and deploying 
scientific applications in high--performance computing environments. Such frameworks provide 
for efficient code reuse, community code development and abstraction of specialized capabilities such 
as adaptive mesh refinement or parallel linear solvers. 

Component specification is obviously an important part  of component frameworks with the specification
providing the definition of the interfaces between components, including for example a description of
the variables and functions both provided by and required by the different components. The choice of 
specification language impacts the scope of capabilities of components which can be implemented and exposed 
as well as the ease of use of 
components by both developers and users. If the component specification is too general it can hinder easy sharing of components, and if the 
specification is too narrow it will reduce the potential functionality of components and thus the application.

%
%\todo{ES: speak about component assembly here.  this is often an
%  explicit step in other frameworks, and is performed implicitly by
%  cactus, which simplifies using thorns significantly.  one
%  disadvantage of the cactus model is that each component can only be
%  activated once -- or maybe that could be added rather simply,
%  without sacrificing automated assembly.  any way, the cactus
%  component specification has to support automated assemly, which
%  means that it also needs to support debugging said assembly (e.g.\
%  by printing the schedule in the beginning).  i'm sure more could be
%  done regarding debugging, e.g.\ listing all variables (we don't do
%  this), or listing all parameters (this happens only upon request,
%  not automatically).  these points should probably go into a section
%  further down, but the automated assembly should be mentioned right
%  here, since it is a key concept of cactus.}
%
%\todo{ES: another important point is that components should be able to
%  be developed very independently, e.g.\ in only loosely-connected
%  collaborations.  this was probably THE success model of cactus in
%  our, let's say, competitive community of numerical relativity.}

This paper describes the current specification of components in the Cactus Framework via the Cactus 
Configuration Language or CCL\@. Cactus is an open--source component framework designed 
for collaborative development of complex codes in high--performance computing environments. 
The largest user base for Cactus  is in the field of numerical relativity where, for example,
over 100 components are now shared among over fifteen different groups through the Einstein Toolkit~\cite{einsteintoolkitweb} (Section~\ref{EinsteinToolkit}). 
In other application areas, Cactus is used by researchers in fields including quantum gravity (Section~\ref{quantumgravity}), computational fluid dynamics, 
coastal modeling and computer science. 

However, as simulation codes grow more complex, for example requiring multi--physics capabilities,
there is now a need to extend or possibly re-architect the CCL to react to new features required by Cactus 
application developers. Further, as the number of Cactus components grow, an increasing problem is 
how to provide user tools for component assembly, application debugging, and verification and validation.
This paper provides a review of the CCL focusing on how it describes the interactions between thorns  and implications
for the development of user tools.

%\todo{ES: mention eclipse, which would aid developers in programming:
%  helping with CCL syntax, understand CCL semantics to be able to
%  refer to CCL-defined entities in the source code, help writing CCL
%  files from entities defined (functions) or used (parameters) in the
%  source code.}

In Section~\ref{cactus} we describe the architecture of the Cactus Framework that particularly
relates to its handling and 
orchestration of components, including the Cactus 
Scheduler, memory allocation, data types provided by Cactus, and existing and planned tools for component management.
In Section~\ref{CCL} we describe the Cactus Thorn 
configuration files using the \ccl, the methods of thorn interaction, and built--in testing options.
In Section~\ref{examples} we examine several different Cactus applications, the WaveToy Demo, a community 
toolkit for quantum gravity, 
and the  Einstein Toolkit, in respect %particularly 
to the dependence among components enforced by the CCL\@.
In Section~\ref{futurework} we describe some ``missing'' features of the CCL that will need to be addressed for future Cactus applications. 

%%%%%%%%%%%%%%%%%%%%%%%%%%%%%%%%%%%%%%%
\section{Cactus}
\label{cactus}

\begin{figure}[t!]
\centering
\includegraphics[width=0.5\linewidth]{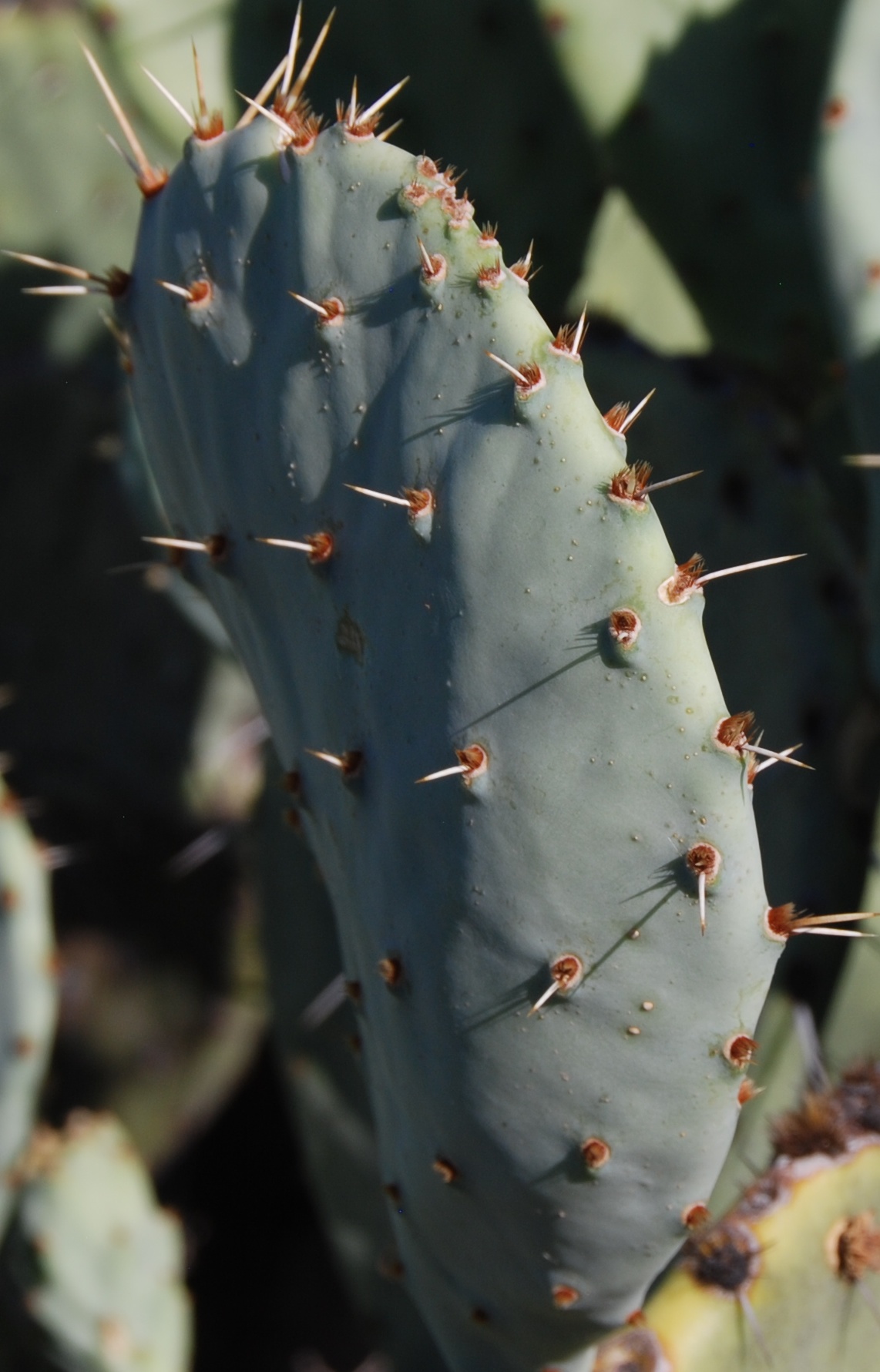}
\label{cactusfig}
\caption{Cactus components are called {\it thorns} and the integrating framework is called the {\it flesh}. The interface between thorns and the flesh is provided by a set of configuration files writing in the Cactus Configuration Language (CCL\@).}
\end{figure}

The Cactus Framework~\cite{Cactusweb,Goodale02a} is an open source, modular, portable programming environment 
for HPC computing. It was designed and written specifically to enable scientists and engineers to 
develop and perform the large--scale simulations needed for modern scientific discoveries across
 a broad range of disciplines. Cactus is well suited for use in large, international research collaborations.

\subsection{Architecture}
\label{cactusarchitecture}

Cactus is a component framework. Its components are called {\it thorns} whereas the 
framework itself is called the {\it flesh} (Figure~\ref{cactusfig}). 
The flesh is the core of Cactus, it provides the APIs for thorns to communicate with each other,
and performs a number of administrative tasks at build--time and run--time. Cactus depends on three
configuration files and two optional files provided by each thorn to direct these tasks and provide inter--thorn APIs. These  files are:
\begin{itemize}
\item{\texttt{interface.ccl}} Defines the thorn {\it interface} and {\it inheritance} along with variables and aliased functions.
\item{\texttt{param.ccl}} Defines parameters which can be specified in a Cactus parameter file and are set at the start of a Cactus run.
\item{\texttt{schedule.ccl}} Defines when and how scheduled functions provided by thorns should be invoked by the Cactus scheduler.
\item{\texttt{configuration.ccl} (optional)} Defines build--time dependencies in terms of provided and required capabilities, e.g. interfaces to Cactus--external libraries.
\item{\texttt{test.ccl} (optional)} Defines how to test a thorn's correctness via regression tests.
\end{itemize}

The flesh is responsible for parsing the configuration files at build-time, generating source code  to 
instantiate the different required thorn variables, parameters and functions,  
as well as checking required thorn dependencies.
% \todo{ES: also scheduling. ??? does it do this at compile time though?}
% FL: Yes, the scheduling tree is generated at compile time
%\todo{ES: the flesh also generates a database that can be accessed at
%  run time -- not all the information is there, but you can ask e.g.\
%  which implementation is provided by which thorn, which
%  variables/groups exist, what parameters there are, etc.  you can't
%  ask about scheduled functions, though, but that's an unfortunate
%  exception -- you can only print the schedule tree\ldots}

At run-time the flesh parses a user provided parameter file which defines which
thorns are required and provides key-value pairs of parameter assignments.~\footnote{Note that
this parameter file is different from the file {\texttt{param.ccl}} which is used to define which
parameters exist, while the former is used to assign values to those parameters at run-time.}
The flesh then activates only the required thorns,
 sets the given parameters, using default values for parameters which are not specified in the parameter file, 
 and creates the schedule of which functions provided by the activated thorns to run at which time. 

% (in an effort to protect users 
%against unexpected effects from other thorns)
%\todo{ES: this is not ``to protect'', and the side-effects are not
%  ``unexpected''.  the other thorns would certainly do something that
%  is not desired.  the other thorns may also provide duplicate
%  functionality, e.g.\ a second driver or a second set of evolved
%  variables. }
%\todo{ES: i would explain this somewhat differently: thorn activation
%  happens at run time instead of at build time, so that changing
%  thorns doesn't require recompiling (a time optimisation), and so
%  that cactus has access to the other thorn's ccl files at build time
%  to ensure consistency.}
%\todo{ES: the flesh also creates the schedule -- automatically, since
%  components assemble automatically!}

The Cactus flesh provides the main iteration loop for simulations (although
this can be overloaded by any thorn)
but does not handle 
memory allocation for variables or parallelization; this is performed by a {\it driver} thorn. 
The 
flesh performs no computation of its own --- this is all done by thorns.
It simply orchestrates the computations defined by the thorns.

\begin{figure}[t!]
\centering
\includegraphics[width=\linewidth]{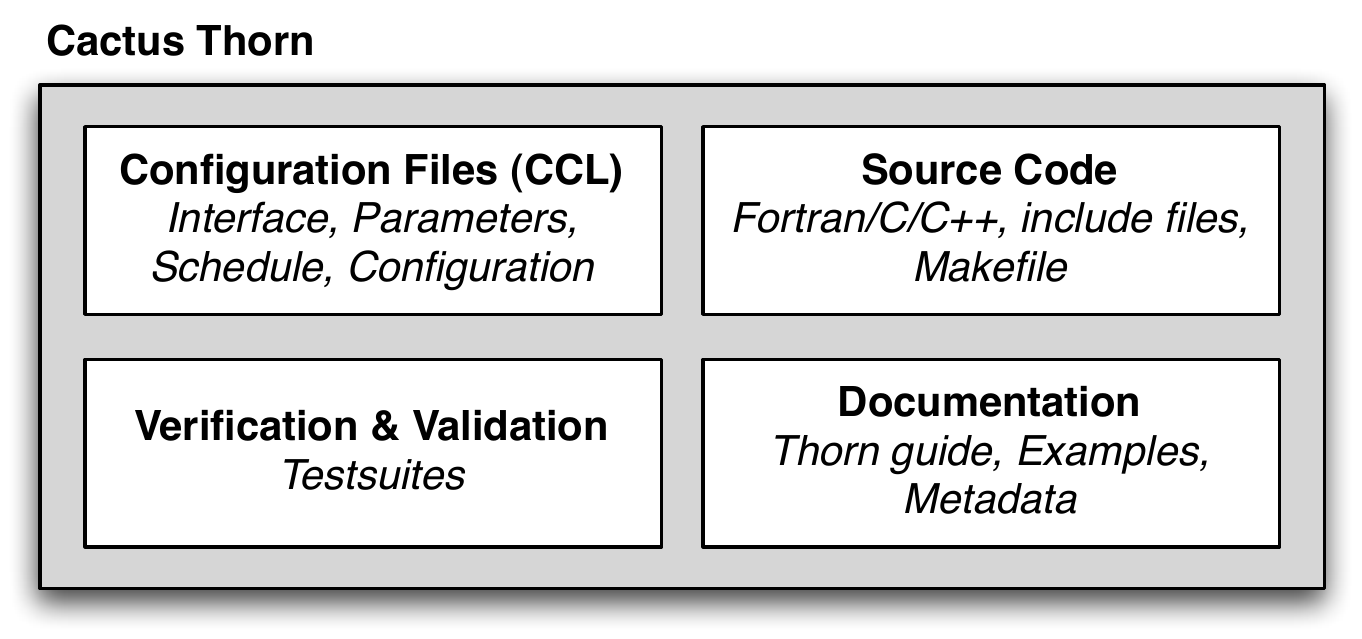}
\label{thornfig}
\caption{Cactus thorns are comprised of source code, documentation, test--suites for regression testing, along with a set of configuration files written in the Cactus Configuration Language (CCL) which define the interface with other thorns and the Cactus flesh.}
\end{figure}

The thorns are the basic modules of Cactus. They are largely independent of each other and
communicate via calls to the Flesh API\@. Thorns are collected into logical groupings called \emph{arrangements},
% , which are
% logical groupings of thorns. 
This is not strictly required, but strongly recommended to aid with 
%thorn
their organization. An important concept %of thorns 
is that of an \emph{interface}.
%which is similar to what other languages might call an \emph{interface}. 
Thorns do not
define relationships with other specific thorns, nor do they communicate directly with other thorns.
Instead they define relationships with an interface, which may be provided by multiple thorns.
This distinction exists so that thorns providing the same interface may be independently
swapped without affecting any other thorns. Interfaces in Cactus are fairly similar to
abstract classes in Java or virtual base classes in C++, with the important distinction that in Cactus
the interface is not explicitly defined anywhere outside of the thorn.
%\todo{ERIC CAN YOU LOOK AT THIS ES: to reduce confusion, i usually say here that an
%  ``implementation'' is what all other languages call an
%  ``interface''.  this corresponds to an ``abstract class'' in Java,
%  or a ``virtual base class'' in C++.  however, different from these
%  languages, there is nothing in cactus which actually defines this
%  interface independent of a thorn -- the implementation is
%  automatically extracted from the public entities defined in a
%  thorn.  this is comparable to a Fortran module, which exports its
%  public entities, whereas a C++ class needs an explicitly written
%  header file.}

This ability to choose among multiple thorns
providing the same interface is important for introducing new capabilities 
in Cactus with minimal changes to other thorns, so that different research groups can implement their own 
particular solver for some problem, yet still take advantage of the large amount of community thorns. For example, 
the original driver thorn for Cactus which handles  domain decomposition and message passing is a unigrid 
driver called {\tt PUGH}. More recently, a driver thorn which implements adaptive mesh refinement (AMR) was developed  
called \texttt{Carpet}~\cite{Schnetter-etal-03b, Schnetter06a, ES-carpetweb}.  Carpet makes it possible for simulations to run with multiple levels of mesh refinement,
which can be used to achieve great accuracy compared to unigrid simulations. Both \texttt{PUGH} and \texttt{Carpet} 
provide the interface \texttt{driver} and application thorns can relatively straightforwardly migrate from unigrid to using the advanced AMR 
thorn. 

Thorns providing the same interface 
 may also be compiled together %at the same time
in the same executable, with the user choosing in the parameter file, at run-time, which implementation to use.
%thorn should provide a particular implementation. 
This allows users to switch
among various thorns without having to recompile Cactus. 

Thorns include a \texttt{doc} directory which provides the documentation for the thorn in 
\LaTeX\ format. This allows users to build one single reference guide
to all thorns via a simple command.

\subsection{Scheduling}

The Cactus flesh provides a rule--based scheduler. Thorn functions can be specified to be called
by the scheduler at different points in the simulation, in standard time bins.  A scheduled routine can be requested to occur before/after other functions in the same timebin.
It is also possible for thorns to define their own \emph{schedule groups}, which may be thought of as a user--defined time bin.
% time bin of its own.
% , or within a new schedule group
% for example.
The 
specification of scheduled functions in thorns is described in Section~\ref{cactusfunctions}. 
At run time, the flesh builds a schedule tree and provides an API that allows this schedule tree to be traversed such that the functions are called in their desired order. Cactus provides the argument lists for calling these scheduled functions, and provides information about which variables need storage allocated and when.

\subsection{Memory Allocation}

Memory allocation for Cactus variables is handled by the driver thorn, using information from the schedule and interface 
configuration files. Memory can be allocated 
for variables throughout the simulation, or allocated only during the execution of a function or schedule group. %group of functions.
This provides a mechanism for reducing and tracking the memory
footprint of a simulation.
Incorrect memory allocation and the use of uninitialized variables can easily lead to bugs in codes which are hard to detect.
% , and different 
Various Cactus thorns provide tools which help locate such errors, for 
example by initializing variables to have a value of {\tt NaN}
\footnote{A full explanation of {\tt NaN} may be found online: \url{http://en.wikipedia.org/wiki/NaN}}
and then checking for these values during the simulation.

\subsection{Data Types}
Cactus defines its own data types for thorns. These data types include standard integer and real types, and a complex number data type.
Supported Cactus data types include \emph{Byte, Int, Real, Complex, String, Keyword and Pointer},
but the use of some of them is restricted (e.g.\ \emph{Keyword}
and \emph{String} to parameters).
An optional trailing number to the type 
can be used to set the size in bytes, where applicable.
The motivation to provide Cactus data types comes from 
the fact that there is not a standard size for data types across all platforms. Providing Cactus-specific
data types allows the framework to maintain an explicit variable size across all platforms, and provides
maximum code portability. In addition it allows users to select the size of these
standard types at build time across all thorns.

\subsection{Tools}

As a distributed software framework, Cactus can make use of %requires 
some additional tools to assemble the code and 
manage the simulations. %Each thorn typically 
Oftentimes each arrangement of thorns resides in its own source control repository, as they are mostly
independent of each other. This leads to a retrieval process that would quickly become
unmanageable for end-users (for example the Einstein Toolkit is comprised of 135 thorns). To facilitate this
process we use a \emph{thornlist} written using the Component Retrieval Language~\cite{TG_CRL}, which allows the maintainers of a 
distributed framework to distribute a single file containing the URLs of the components and the
desired directory structure. This file can then be processed by a program such as our own 
\texttt{GetComponents} script, and the entire retrieval process becomes automated.

In addition to the complex retrieval process,
compiling Cactus and managing simulations
can be a difficult task, especially for new users. There are a large number of options that may
be required for a successful compilation, and these will vary across various architectures. To assist
with this process a tool called the \emph{Simulation
  Factory}~\cite{ES-simfactoryweb, CBHPC_SimFactory} was 
developed. Simulation Factory provides a central means of control for managing access to different 
resources, configuring and building the Cactus codebase, and also managing the simulations 
created using Cactus. Simulation Factory uses a database known as the \emph{Machine Database}, 
which allows Simulation Factory to be resource agnostic, allowing it to run consistently across any 
pre-configured HPC resource.

%%%%%%%%%%%%%%%%%%%%%%%%%%%%%%%%%%%%%%%%
\section{Cactus Configuration Language}
\label{CCL}

The Cactus Configuration Language (CCL) was provided with the first Cactus 4.0 release in 1999. 
The language has evolved since then with the addition of function aliasing (Section~\ref{cactusfunctions}) and the configuration CCL file (Section~\ref{cactusarchitecture}), along with a small number of minor changes.
The well designed initial capabilities and ensuing stability of the CCL is one feature of Cactus which 
has led to its success across different scientific fields and its ability to enable the growth of application communities.

In this section we outline the structure of the \ccl\ and provide syntax definitions for many of the
elements of CCL\@. A complete specification and discussion of the language may be found in the 
Cactus User's Guide\footnote{\url{http://cactuscode.org/documentation/UsersGuide.pdf}}.

\subsection{Thorn Configuration}
\subsubsection{Groups}

Cactus variables are placed in variable groups with homogeneous attributes,
where the attributes describe properties such as the data type, variable group
type, rank, dimensions, and number of time levels. 
%\todo{ES: don't mention concepts here that have not been introduced --
%``group type'', ``ghost size'', etc.\ are not understandable.}
%FL: I removed some of them, but items like 'group type' are generic enough to be
% understood to a basic level until they are explained in detail
Many Cactus functions operate on groups of variables, for example storage allocation, sychronization between processors, and output
functions.
%\todo{ES: also declaring variables (can only be done in groups), public/private properties.}
% FL: It is possible to declare variables without explicitly declaring a group. Cactus
% does create a group with the same name as the variable then, but that's an internal
% detail
%\todo{ES: maybe mention that groups cannot contain groups.}
For example, a vector field containing individual variables for fluid flow in different directions would typically include all the vector components in 
a single variable group. 
By default, all variable groups are private,
however the {\tt public} keyword can be used to change the access level for each subsequent variable group in the ccl file.

\subsubsection{Functions}
\label{cactusfunctions}
Cactus provides two types of functions, \emph{scheduled} and \emph{aliased}. Scheduled
functions are declared in the \texttt{schedule.ccl} file and are defined to be called at certain stages
in the Cactus simulation by prescribing a {\it time bin}, a specific time during a simulation,
in which to run. Standard Cactus time bins are defined 
which are invoked in a well defined order, and a list of 
 the most important Cactus standard time bins is provided in Figure~\ref{CactusTime Bins}. 

\begin{figure}
{\small
\begin{tabular}{|l|p{49 mm}|}
\hline
{\bf Schedule Bin} & {\bf Description} \\
\hline \hline
\texttt{CCTK\_STARTUP} &
For routines which need to be run before the grid hierarchy is set up, for example, for function
registration.\\
\hline
\texttt{CCTK\_PARAMCHECK} &
For routines that check parameter combinations for potential errors. Routines registered here
only have access to the grid size and the parameters.\\
\hline
\texttt{CCTK\_INITIAL} &
For routines which generate initial data.\\
\hline
\texttt{CCTK\_PRESTEP} &
Tasks performed before the main evolution step.\\
\hline
\texttt{CCTK\_EVOL} &
The evolution step.\\
\hline
\texttt{CCTK\_POSTSTEP} &
Tasks performed after the evolution step.\\
\hline
\texttt{CCTK\_ANALYSIS} &
Routines which can analyze data at each iteration. This time bin is special in that ANALYSIS routines are only called if output from the routine is requested, e.g. in the parameter file\\
\hline
\end{tabular}
}
\caption{Scheduled functions in Cactus can be assigned to run in standard time bins, the most important of which are described in this table.}
\label{CactusTime Bins}
\end{figure}
%\todo{ES: this speacialty of analysis is a nice feature -- but it
%  isn't important.  it either needs to be explained better (what is
%  ``requesting output from the routine''?), or left out.}
% FL: I hope that's now explained well enough, for the limited scope of the paper

Additionally, 
thorn developers can define their own time bins or schedule groups.
It is possible to specify the order in
which two scheduled functions are called,
as well as simple conditionals and loops.
Memory allocation of Cactus variables can be restricted to only the time of execution of a
certain function.
Figure~\ref{cactusscheduling} shows a subset of the syntax which is used to define a scheduled function.

\begin{figure}[h!]
\centering
{\small
\begin{Verbatim}[frame=single, framerule=0.3mm]
SCHEDULE [GROUP] 
  <function|schedule group name> 
  AT|IN <schedule bin|group name> 
  [WHILE <variable>] [IF <variable>] 
  [BEFORE|AFTER <item>|(<item> <item> ...)]*
{
  [STORAGE: <group >,<group >...]
  [SYNC:    <group >,<group >...] 
} "Description of function or schedule group"
\end{Verbatim}
}
\caption{Subset of the syntax for declaring scheduled functions or schedule groups of functions.
A function can be scheduled at a certain time bin or in a schedule group.
It can be called while or if a condition is fulfilled. Functions or schedule groups can be
scheduled before or after other functions or schedule groups, within the same time bin or schedule group.
Storage for Cactus variables might only be allocated for a certain function or schedule group,
to save overall memory. Variables distributed over multiple processes can be automatically synchronized after a certain function or schedule group, if specified in the ccl file.
}
\label{cactusscheduling}
\end{figure}

Aliased functions are functions that can be shared between thorns. They are declared in the 
\texttt{interface.ccl} file and may be called by a thorn at any point during the simulation.
In order to call an aliased function it is not important to know the programming language used
for its implementation. The Cactus API takes care of possibly necessary conversions.

\subsubsection{Variables}
\label{sec:variables}
\emph{Grid variables} are Cactus variables that are passed between thorns by the flesh, and are declared in the \texttt{interface.ccl} file. 
They are generally collected into \emph{variable groups} of the same data type.
There are three types of variable groups: \emph{grid functions}, \emph{arrays}, and \emph{scalars}.
\emph{Grid functions} (GFs), the most common variable group type, are arrays with a specific size set by the
parameter file, which are distributed across processors. All GFs must have the same array size, typically defining the shape and size of the computational domain.
\emph{Arrays} are a more general form of GFs in that each array group may have a distinct size which
can be given by Cactus parameters. 
\emph{Scalars} are single variables of a given basic type, much like rank-zero arrays.
Cactus variables can specify a number of timelevels, which means a certain number
of copies of this variable for use in time--evolution schemes where data at a past time is needed to
calculate the new data at a later time.
Part of the syntax for declaring a variable group of variables is shown in Figure~\ref{cactusvar}.

\begin{figure}[h]
\centering
{\small
\begin{Verbatim}[frame=single, framerule=0.3mm]
<data_type> <group_name> 
    [TYPE=<group_type>] 
    [SIZE=<size in each direction>] 
    [TIMELEVELS=<num>]
[{ 
 [ <variable_name>[,]<variable_name>
   <variable_name> ] 
} ["<group_description>"] ]
\end{Verbatim}
}
\caption{Part of the syntax for declaring Cactus variables. Cactus variables have to be one
of the data types Cactus defines and are part of a variable group. They can have different
Cactus variable types, sizes, and number of time levels.
Each variable group needs to have a human--readable description.}
\label{cactusvar}
\end{figure}

\subsubsection{Parameters}
\emph{Parameters} are used to specify the runtime behavior of Cactus and are defined in the
\texttt{param.ccl} file. They have a specific data type and scope, a range of allowed values, and a default value.
Once parameters have been set, they cannot be modified unless specifically declared to be 
\emph{steerable}, in which case they may be dynamically changed throughout the simulation. 
The allowed datatypes for parameters are \emph{Int}, \emph{Real}, \emph{Keyword}, \emph{Boolean}, and \emph{String}. Thorns can use and extend parameters of other thorns.
The syntax for declaring Cactus parameters is shown in Figure~\ref{cactuspar}.

\begin{figure}[h]
\centering
{\small
\begin{Verbatim}[frame=single, framerule=0.3mm]
[EXTENDS|USES] <parameter_type>
  <parameter name> "<parameter description>" 
{
  <PARAMETER_RANGES> :: "Range description"
} <default value>
\end{Verbatim}
}
\caption{Syntax for declaring Cactus parameters. Thorns might use or extend parameters of
other thorns, and define their own. A parameter needs to have a data type. A human--readable description needs to be given, as well as an allowed range
with a description for the range and a default value within that range.}
\label{cactuspar}
\end{figure}

\subsubsection{Include Files}
\label{includes}
Header files can be shared between thorns if specified in the \texttt{interface.ccl} file. It is
not only possible to share a single include file, but also to concatenate multiple include
files (also from multiple thorns), and use them like a single include file.
During the build process, Cactus copies all of the source files located in each 
thorn's \texttt{include} directory to a central location from which they may be accessed by any
other thorn using one of two methods shown in Figure~\ref{cactusinclude}. \texttt{USES INCLUDE} requests
an include file from another thorn, and 
\texttt{INCLUDE} adds the code in \texttt{\emph{<file\_to\_include>}}
to \texttt{\emph{<file\_name>}}.

\begin{figure}[h]
\centering
{\small
\begin{Verbatim}[frame=single, framerule=0.3mm]
USES INCLUDE: <file_name>
INCLUDE[S]: <file_to_include> IN <file_name>
\end{Verbatim}
}
\caption{Syntax for using include files in Cactus. Thorns might provide a specific header file
to another thorn (the first example), or might provide one part of a concatenation of multiple header files, possibly from multiple thorns (the latter example).}
\label{cactusinclude}
\end{figure}

\subsection{Thorn Interaction}
\subsubsection{Scope}
Cactus provides different levels of access for variables and parameters. 
Variables can be defined as \emph{public} or \emph{private}.
\emph{Public} variables can be inherited by a thorn when that thorn inherits an interface. 
Thorn inheritance will be described in greater detail below. 
%\emph{Protected} variables are shared with other thorns when the thorns are declared to be friends. 
\emph{Private} variables can only be seen by the thorn which defines them.
%\todo{ES: leave out ``protected'' and ``friends'' here.  these
%  concepts are too complex to be described here, and they are outdated.}

Similarly, parameters may be defined as \emph{restricted} or \emph{private}.
%\emph{Global} parameters are visible to all other thorns.
\emph{Restricted} parameters are available to thorns which request access.
\emph{Private} parameters, like variables, are only visible to the thorn which defines them.
The
access levels here only specify if those parameters are directly accessible in the source code; it is possible to access information about any parameter through Cactus API functions regardless of 
the parameter scope defined in the {\tt param.ccl} file. 
%\todo{ES: leave out ``global'' here, it is outdated.}

\subsubsection{Inheritance}
Cactus provides an inheritance mechanism similar to Java's abstract classes. It allows thorns
to gain access to variables provided elsewhere by \emph{inheriting} from the interface. A key point here
is that the thorns are not inheriting from other specific thorns; any number of thorns may declare
themselves as implementing an interface. These thorns may all be
compiled together, allowing the user to decide at run-time which thorn should be used.
The interface is only specified by the thorns implementing it. This means
that thorns declaring the same interface-name need to have an identical interface,
which is checked by Cactus.

Cactus also provides \emph{capabilities} which may be declared in the \texttt{configuration.ccl} file.
Capabilities differ slightly from interfaces in that while any number of thorns providing the
same interface may be compiled together, only one thorn providing a capability may be
compiled into a specific configuration. In this sense, while interfaces define run--time
dependencies, capabilities define build--time dependencies.
This can be useful for providing external libraries or 
functions which are too complex for aliasing. Also, capabilities play a role in 
configuring thorns and external libraries since they interact with the build system of Cactus.

Many design decisions are based on the distinction between interfaces and capabilities.
For example,  the 
concept of capabilities is important for application performance
-- knowing an inter-thorn relationship at build time allows
optimizations to be included that are not possible at run time.  

The syntax for declaring and requiring a capability is shown in
Figure~\ref{cactus_capability_provides}.

\begin{figure}[h]
\centering
{\small
\begin{Verbatim}[frame=single, framerule=0.3mm]
PROVIDES <Capability> 
{
  SCRIPT <Configuration script> 
  LANG <Language> 
}

REQUIRES <Capability>
\end{Verbatim}
}
\caption{Part of the syntax for declaring and requiring capabilities in Cactus. Capabilities can be
required and provided by thorns. If a thorn provides a capability it interacts with the makesystem
through the output of a script which needs to be specified in the ccl file, as well as it's
programming language to be able to call it correctly.}
\label{cactus_capability_provides}
%\todo{ES: need more detail here, explaining the syntax.}
\end{figure}

%\begin{figure}[h]
%\centering
%{\small
%\begin{Verbatim}[frame=single, framerule=0.3mm]
%
%OPTIONAL <Capability> 
%{
%  DEFINE <macro>
%}
%\end{Verbatim}
%}
%\caption{Syntax for requesting capabilities in Cactus}
%\label{cactus_capability_requires}
%\todo{ES: is OPTIONAL implemented?}
%\end{figure}

The \texttt{interface.ccl} file also provides a low-level \emph{include} mechanism, described in 
Section~\ref{includes}, similar to that
found in C/C++. Thorns may request access to any include file within the Cactus source tree
without specifying which thorn or interface should provide it. This is used primarily for optimization
reasons as the compiler can then replace inline functions, and in some cases for providing access
to external libraries such as HDF5.

\subsection{Testing}
It is strongly recommended, although not required, that thorns come with one or more test suites.
These consist of sample parameter files and the expected output for those parameters. These files
should be located within the \texttt{test} directory in the thorn, so that the test suites may be run
using \texttt{gmake \emph{<configuration>}-testsuite}. These test suites serve the dual purposes
of regression and portability testing.

%\subsection{Design Points}

%What we were trying to achieve

%\subsection{Configuration Files}

%Describe the different CCL files at the moment (not the language in them, just what each one does)

%\subsection{CCL Specification}

%The actual specification, maybe need subsubsections? Need to write as BNF?

%%%%%%%%%%%%%%%%%%%%%%%%%%%%%%%%%%%%

%%%%%%%%%%%%%%%%%%%%%%%%%%%%%%%%%%%
\section{Examples}
\label{examples}

In this section we show some examples of the dependencies among Cactus thorns which are generated by the CCL files 
for different applications: a simple example application for the scalar wave equation 
with a minimal set of thorns; a small community toolkit for quantum gravity; and a large community toolkit for numerical relativity. The interest on thorn dependencies arises for two core reasons:

\begin{enumerate}

\item Cactus is particularly targeted at enabling communities to generate shared toolkits for solving a variety of problems in a particular field. The standard computational toolkit which is distributed with Cactus is further used by many different applications. Thorn dependencies and interfaces thus need to be carefully thought out and periodically revisited to make sure that the {\it plug-and-play} aim of Cactus, where different thorns can provide the same functionality, is achieved with interfaces which are as simple, flexible and general as possible. This design usually involves a delicate balance, taking into account   the speed of implementation, complexity of the interface etc.

\item Long time Cactus users work with standard thorn lists which are built up from experience and shared with collaborators. These thorn lists are amended as new thorns become available or are no longer used, and can contain several hundred thorns. For new users in particular, there is an 
increasing issue with providing a procedure for users to select the appropriate set of thorns for their application, and to understand the capabilities of different thorns. One big simplification which could be made would be to reduce the number of thorns in thorn lists by removing thorns which depend on others and could be automatically added. Ideally, a tool would be built which would allow a user to start from an abstract description of their problem and automatically select appropriate thorns, for example {\it Evolving Gaussian initial data using the 3-D scalar wave equation and outputting 3D data}, or {\it Evolving two black holes using Einstein's equations and calculating gravitational waveforms}. The question is then whether there is currently enough information in the CCL files to achieve this, or how additional information could be provided.

\end{enumerate}

In this section, we use the dependencies among the sets of thorns described in the CCL files for these three example applications to view 
the complete set of thorn dependencies and to investigate how the thorn set could potentially be generated from an initial minimal set of 
thorns. The dependencies used for the figures are taken from a file generated during the Cactus build process which contains a complete database 
of the contents of the different thorn configuration files. 

A Perl script is used to parse this database and generate a file in \texttt{dot} format, which
can then be processed by a program like \texttt{graphviz}~\cite{graphvizweb} and turned into a directed graph like that
in Figure~\ref{WaveDemo_cactus}. This graph shows five different types of dependencies. 
Inheritance is denoted by a regular arrow, dependencies due to a required function are denoted by 
an arrow with a square head, direct thorn dependencies are denoted by a dotted arrow, shared
variable dependencies are denoted by an arrow with a circular head, and dependencies due to a
required capability are denoted by an arrow with a diamond head. There are also shaded and 
unshaded thorns, the distinction being that the shaded thorns have no other thorns depending on them.

This Perl script does not show the dependencies generated by a single thorn, so we also use a set of two Python
scripts, the first of which parses the actual CCL files and generates an XML file containing all of the dependencies.
This file can then be queried by the second script, which will search for a single thorn and find
all thorns upon which the query depends. It will also output a graph in \texttt{dot} format, as seen in
Figure~\ref{WaveDemo_eric}. The second script will also allow users to choose between
alternate implementations of the same interface (e.g.\ \texttt{PUGH} or \texttt{carpet}). The motivation
here is that this script should allow the user to generate a complete thornlist that could then be used
to build a simulation.

\subsection{Simple Example: Scalar Waves}

\begin{figure*}[t!]
\centering
\includegraphics[width=\linewidth]{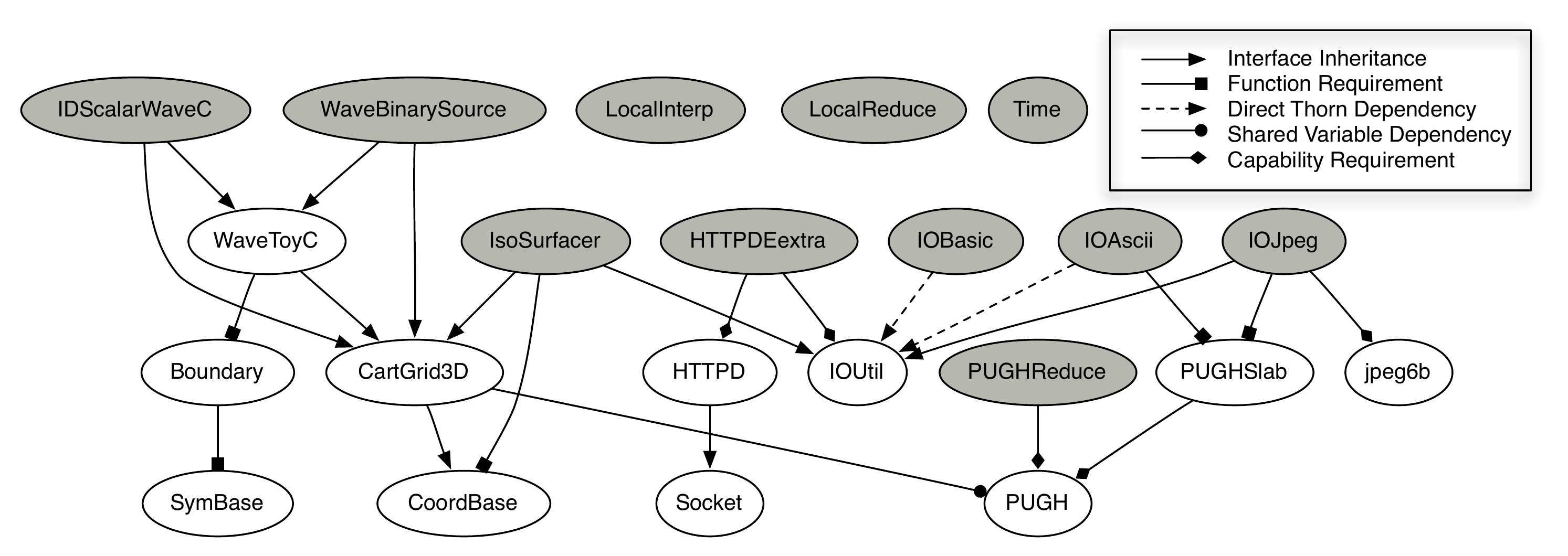}
\caption{\label{WaveDemo_cactus}
Dependency graph for complete set of thorns in the simple example application {\tt 
WaveToy Demo}. The shaded items indicate that the thorns are `leaves' and have no thorns 
depending on them.}
\end{figure*}

The set of Cactus thorns to solve the 3-D scalar wave equation (WaveToy Demo) was developed as a pedagogical example 
for understanding Cactus, and as a simple and well understood test case for new developments. These
thorns solve the hyperbolic wave equation in 3D Cartesian coordinates with different boundary conditions for a chosen set of initial data and include 
different output formats and a web interface. This example is described 
on the Cactus web pages~\cite{Cactusweb}, which also provide a thorn list with information about the 22 thorns that are used. 
The example application includes two initial data thorns which specify the initial scalar field and sources ({\tt idscalarwavec} and {\tt wavebinarysource}), 
a scalar field evolver ({\tt wavetoyc}) along with additional 
thorns from the standard Cactus Computational Toolkit. 
The example uses the unigrid driver {\tt pugh} with associated thorns {\tt pughslab} for 
hyperslabbing and {\tt pughreduce} which provides a set of standard reduction operations that can calculate for example 
the maximum value or L2 norm over the grid for any grid variable.

\begin{figure}[h]
\centering
\includegraphics[width=0.7\linewidth]{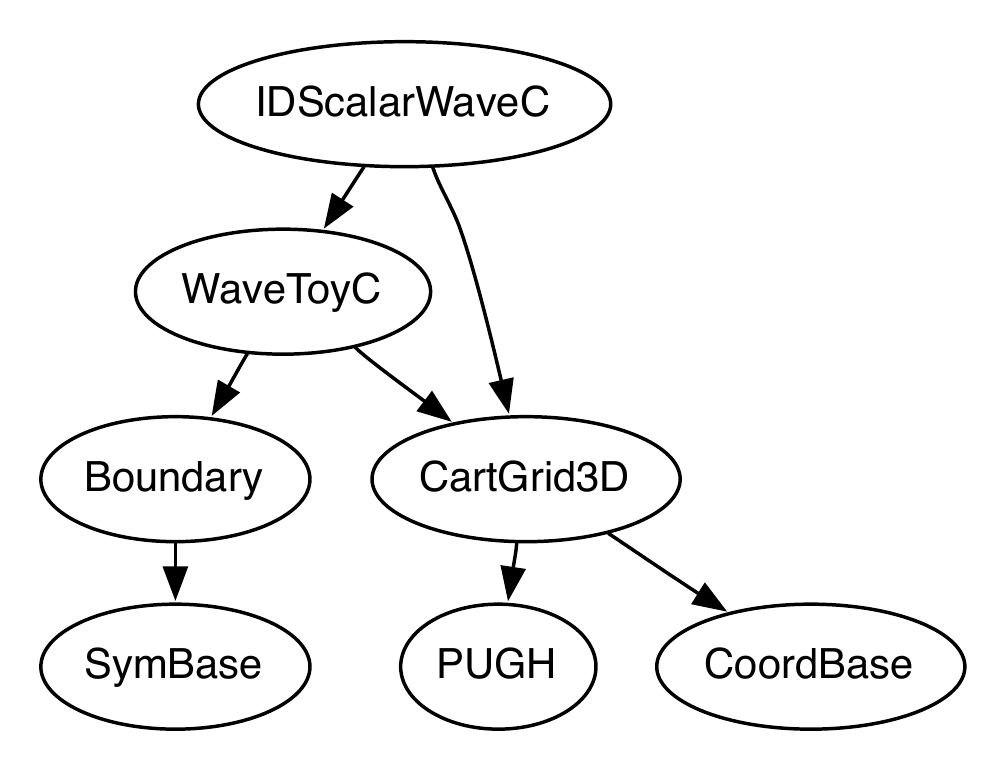}
\label{WaveDemo_eric}
\caption{Dependency graph for the WaveToy Demo thornlist. This graph is generated using dependencies of thorn IDScalarWaveC which defines initial data for 
the fields evolved by the scalar wave equation.}
\end{figure}

 A complete set of dependencies between 
these thorns as specified in the CCL files is shown in Figure~\ref{WaveDemo_cactus}. In this diagram
we can see for example the central nature of the {\tt ioutil} thorn which provides functionality 
that can be used by thorns implementing different I/O methods, for example providing a parameter
which sets when data for all I/O methods should be output and the directory in which to write data.

The dependency diagram also shows that any method to automatically generate this set of thorns 
using dependency information would need 11 thorns specified as a starting point, these are the shaded
thorns in the diagram. For example, if we simply started from the thorn that specifies the 
initial scalar field ({\tt idscalarwavec}) as shown in Figure~\ref{WaveDemo_eric}, which could be 
the obvious starting point for 
a user who knows they want to evolve a particular scalar field then working only with dependencies 
would result in a set of thorns without using any coordinate time ({\tt time}), any I/O, or the possibility to include scalar source terms.

Adding additional metadata to thorns is one mechanism to supplement the current CCL information to enable the generation of thorn lists for a particular 
application. For example, explicitly tagging thorns as providing I/O methods would allow these
thorns to be automatically added or to be selected by a user. In other cases, these diagrams show 
that additional interfaces or dependencies may need to be added. In Figure~\ref{WaveDemo_eric} 
attention needs to be given to the compile time dependencies that would include thorns {\tt time} 
(which should in fact be inherited by the evolution thorn) and {\tt PUGHReduce} and {\tt localreduce}.

\subsection{Small Community Code: The CausalSets Toolkit}
\label{quantumgravity}

The CausalSets Toolkit is an example of a small community codebase, which
implements a wide variety of computations in discrete quantum gravity, in
particular with regard to Causal Set Theory \cite{causets}.
The toolkit is
based upon two major components. One is a MonteCarlo arrangement, which
provides a generic API for providing parallel random numbers, i.e.\ pseudo
random numbers which are independent on all processes.  A second is a
CausetBase API, provided by the BinaryCauset thorn, which abstracts the
mathematical notion of a causal set (a locally finite partially ordered set
\cite{causets}), providing myriad routines for working with such objects.

One of the
challenges in supporting computations in Causal Set Theory is that 
% one would like to perform 
there is
not a single sort of computation, such as finding approximate solutions to
PDEs by finite difference or spectral methods, which one would like to
perform.
Instead a physicist will ask many different sorts of questions about the
behavior of discrete partial orders.  A given computation will share aspects
with others, but the overall structure may differ considerably.
% , which may share some aspects with others
% discrete quantum gravity is th
Furthermore the community is in general not terribly experienced with large
scale computation, and thus benefits from software which insulates the
physicist from many
complications of parallel computing.
The component based approach provided by the Cactus Framework is well suited
to address both of these challenges, by allowing the physicist to mix and
match individual components to build up the particular computation desired,
working with familiar abstract mathematical concepts, rather than having to
work directly with source code.  Additionally the components are designed
to run readily on large scale hybrid architectures, without the user needing detailed
knowledge of how the computation is implemented.

\begin{figure}[h]
\centering
\includegraphics[width=\linewidth]{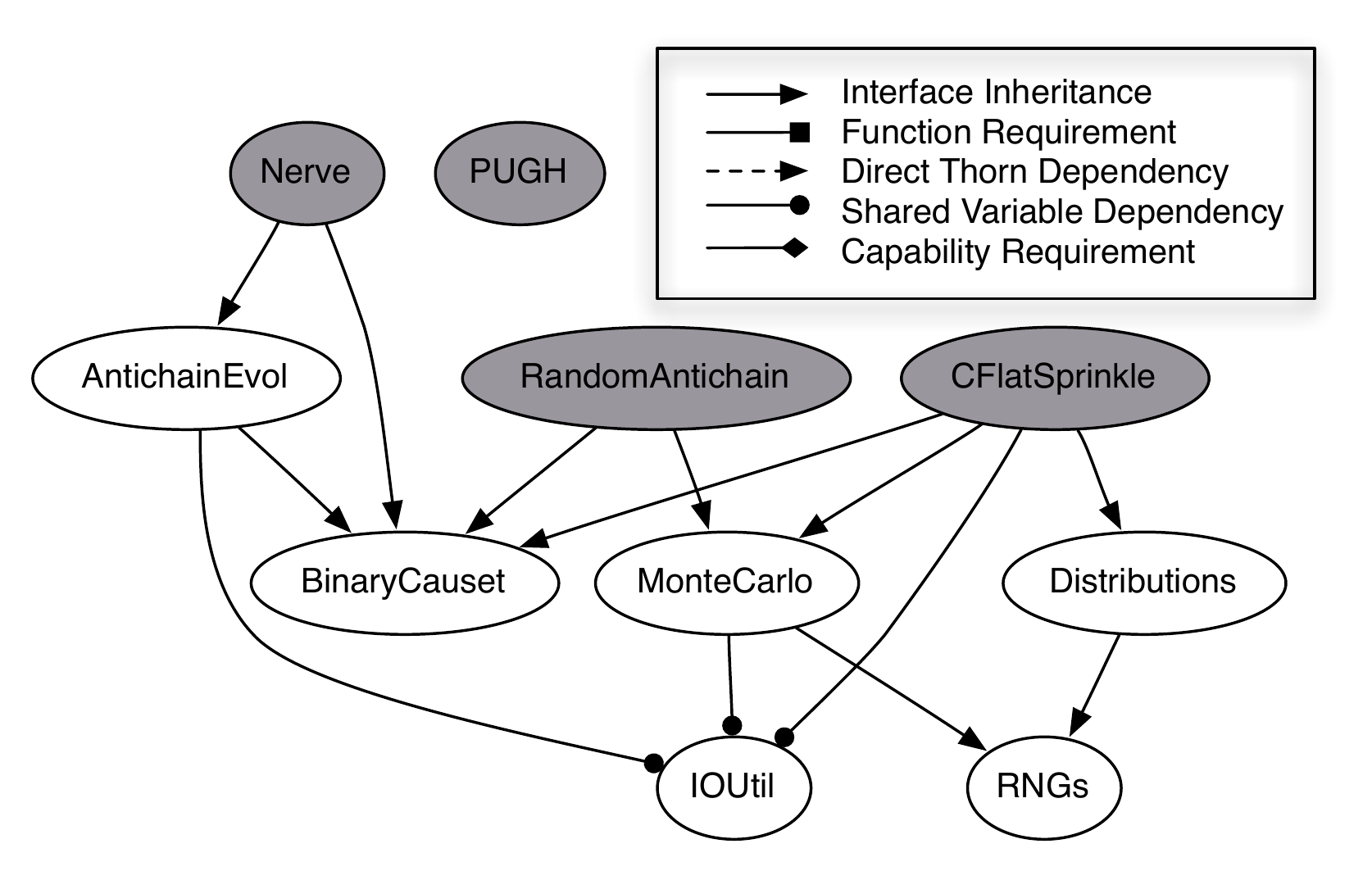}
\caption{Dependency graph for a sample computation in Causal Set Quantum
  Gravity.  The computation is described in detail in \cite{numhomology}.}
\label{CausalSetsFull}
\end{figure}

The dependency diagram for a collection of thorns which implements a sample
computation is shown in Figure \ref{CausalSetsFull}.  This is a computation
of spatial homology of a sprinkled causal set, as described in \cite{numhomology}.
Here the BinaryCauset
thorn implements the core CausetBase API, which provides the causal set along
with a high level abstract interface to it.  The MonteCarlo thorn provides
parallel random numbers to CFlatSprinkle, which generates a random causal
set, and RandomAntichain, which selects a random antichain within the causal
set provided by CFlatSprinkle.  
The MonteCarlo arrangement gets the actual pseudorandom numbers from thorn
RNGs, and also provides a thorn Distributions to provide samples from a
variety of distributions, such as Poisson and Gaussian.
AntichainEvol provides a sequence of
`thickened antichains', which are then read by the Nerve thorn, which
computes a nerve simplicial complex from each thickened antichain.  The
homology groups of these
simplicial complexes are then computed by a separate standalone homology
package {\tt chomp} \cite{chomp}.
The whole computation relies on PUGH as a standard Cactus driver, and uses
Cactus' IOUtil to provide metadata for IO routines.

% \begin{figure}[h]
% \centering
% \includegraphics[width=0.7\linewidth]{CausalSets_eric}
% \label{CausalSets_eric}
% \caption{}
% \end{figure}

\subsection{Large Community Code: The Einstein Toolkit}
\label{EinsteinToolkit}

The Einstein Toolkit~\cite{einsteintoolkitweb} is an open, community developed software infrastructure for relativistic astrophysics. The Einstein Toolkit is a 
collection of software components and tools for simulating and analyzing general relativistic astrophysical systems that builds on numerous software efforts in the numerical relativity community. 
The Cactus Framework is used as the underlying computational infrastructure providing large-scale parallelization, 
general computational components, and a model for collaborative, portable code development. The toolkit includes modules 
to build complete codes for simulating black hole spacetimes as well as systems governed by relativistic hydrodynamics. 
Current development in the consortium is targeted at providing additional infrastructure for general relativistic magnetohydrodynamics.

The Einstein Toolkit uses a distributed software model and its different modules are developed, 
distributed, and supported either by the core team of Einstein Toolkit Maintainers, or by individual 
groups. When modules are provided by external groups, the Einstein Toolkit Maintainers provide 
quality control for modules for inclusion in the toolkit and help coordinate support. 

With such a large set of components and a distributed team of developers, implementing appropriate standards are crucial to maintain coherence across 
the code base, and to enable future development. This is achieved in some part by defining {\it base} 
thorns that act to define application specific standards, providing default variables, parameters, 
functions and schedule bins that are common across an application. For example, in the Einstein 
Toolkit application specific base thorns include {\tt ADMBase} (for the vacuum spacetimes), {\tt 
HydroBase} (for matter spacetimes) and {\tt EOSBase} (for equations of state)~\cite{1456196}.

\begin{figure*}[t!]
\centering
\includegraphics[width=\linewidth]{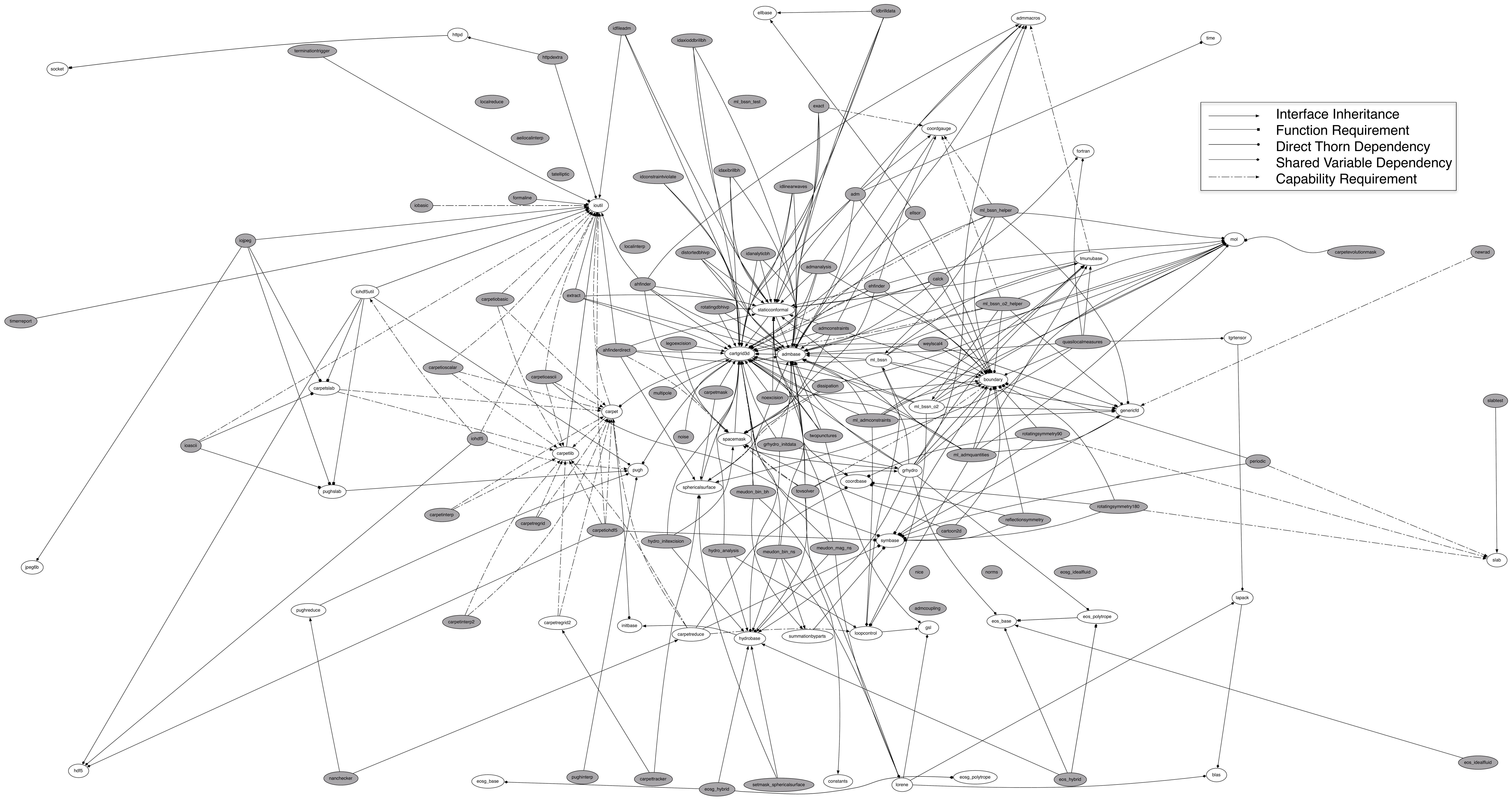}
\label{EinsteinToolkitCactus}
\caption{Complete dependency graph for the {\tt 135} thorns of the EinsteinToolkit (\url{http://www.einsteintoolkit.org})}
\end{figure*}

Figure~\ref{EinsteinToolkitCactus} shows the complete dependency graph for the Einstein Toolkit, 
which is so extensive that it isn't possible to examine in detail in print\footnote{Note that if viewing this paper as a PDF document it is possible to zoom in to see features in detail.}; however, we include the 
graph here to illustrate its complexity. Of the 135 thorns, 9 have no dependency on other thorns, and 78 thorns (including these independent 
thorns) are needed as the starting point to generate the whole toolkit using CCL dependency 
information. The clusters of dependencies for {\tt ADMBase}, {\tt HydroBase} and {\tt EOSBase} are 
apparent in the diagram. 

The Einstein Toolkit dependency diagram also shows a number of direct thorn dependencies, indicated by the black dotted lines. This means that
thorns depend not on an interface but on a specific thorn. In some cases this is due to missing general interfaces such as appropriate aliased functions which either need to be carefully designed or perhaps have simply not been added where they should have been. A large number of these 
direct dependencies are associated with the Carpet adaptive mesh refinement set of thorns where the nature of the driver thorn typically enforces 
a direct dependency for example for associated I/O or reduction operations. The need to support direct dependency on thorns was one reason why the
{\tt configuration.ccl} file was introduced as an extension to the original CCL.

\begin{figure}[h]
\centering
\includegraphics[width=0.7\linewidth]{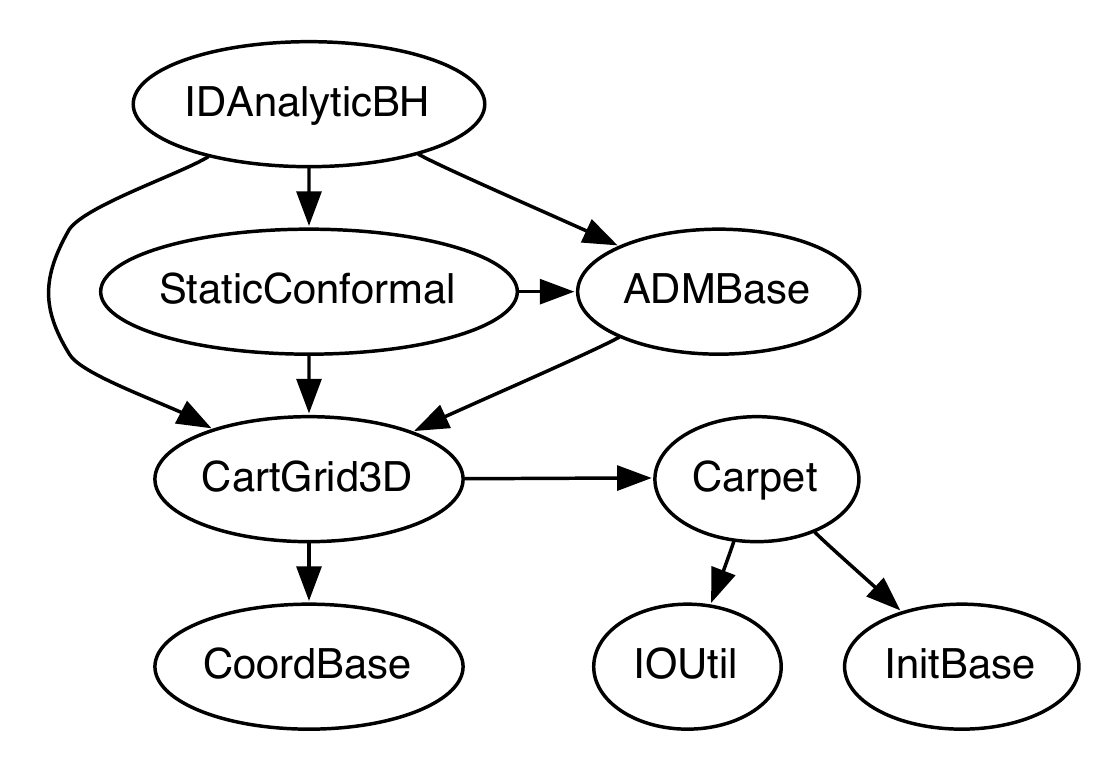}
\label{IDAnalyticBH}
\caption{Dependency graph for the Einstein Toolkit starting from the IDAnalyticBH thorn. For this 
graph the thorn Carpet was chosen to provide the \emph{driver} interface, however PUGH could 
have been used instead.}
\end{figure}

Figure~\ref{IDAnalyticBH} shows an example of the direct dependencies for an initial data thorn in the Einstein Toolkit. The thorn {\tt IDAnalyticBH} 
provides initial data for several different black hole spacetimes with analytic solutions. Starting from this thorn, only seven other thorns are picked up 
directly with dependency information. Given that most production runs for numerical relativity simulations include of order 100 thorns, it is clear that
automatically generating appropriate thorn lists will require additional metadata and physics insight.

\section{Future Work}
\label{futurework}
The original \ccl\ was released as part of the Cactus 4.0b distribution in 1999 and has since that time been  extended in different ways as new features were required. 
Despite serving the Cactus user community well since this time, it is clearly time to reexamine the requirements for the CCL in the 
light of current and future needs and to take into account new technologies and possibilities. 
 In this section we describe new features required in the CCL and their motivation. 
 
 Cactus (and the set of thorns in the Cactus Computational Toolkit) 
currently best supports finite difference, finite volume, or finite element methods implemented on structured grids. Extensions to the CCL 
are required to support meshless methods (e.g.\ particle methods such as smoothed particle hydrodynamics or particle-in-cell, used for 
example in many astrophysics codes) and unstructured meshes where additional connectivity information is required to specify how 
grid points are connected (e.g.\ unstructured grids are important for example in coastal modeling to resolve the fine details of the coastline). 
Implementing both these features in Cactus requires developing appropriate parallel driver and associated infrastructure thorns 
in addition to changes to the CCL\@. 
 
Cactus currently operates with a single computational grid so that all physical models need to run on a single domain. Comprehensive multiphysics support is needed 
where different physical models can be configured and run on different domains, for example for coupling together wind and current models 
in coastal science, or modeling different physical components of a relativistic star. 
 
Constants (e.g.\ $\pi$ or the solar mass) are commonly used in
  scientific codes. Currently in Cactus constants are handled via
  include files, for example the Einstein Toolkit contains a thorn
  which provides commonly used astrophysical constants in an include
  file.  These constants are then only available in source code and
  not in CCL files.  A preferable approach would be to define such
  constants directly as part of the CCL specification.

Similar to constants, the CCL needs to support enumerations and
  user-defined structures, so that e.g.\ a hydrodynamical state vector
  consisting of density, velocity, and temperature can be handled as a
  combined entity instead of as a set of five separate variables.  This
  should include the ability to handle vectors and tensors in a
  natural manner, a feature that is missing in many computer
  languages, but which is nevertheless important in physics
  simulations.  Tensor support would need to include support for
  symmetries (so that e.g.\ only 6 out of 9 components of the stress tensor
  are stored).
  In implementing this, it is important that the abstract
  specification of data types is decoupled from the decision of how to
  lay them out in memory, which needs to be left to the driver to
  ensure the highest possible performance on modern architectures that
  may offer vectorization and deep cache hierarchies.

While Cactus, through the CCL, contains information on how thorns fit together computationally the CCL does not contain information on the scientific content of the thorns. This issue needs some attention as the number of thorns in particular domains grows and models become more complex. 
Options to handle this could include extending the CCL, or adding descriptive metadata separate to the CCL, or by investigating whether enough information can be provided from the CCL and {\it base} thorns for a particular application. Such additional information is important, for example, to be able to automatically construct appropriate thornlists for a particular physical model.

A further issue related to the growth in both the number of thorns and the complexity of applications is constructing and editing CCL files. CCL files for some thorns are now very long and complex and difficult to read and comprehend. This issue could be addressed by restructuring the CCL itself or by providing intuitive and  flexible higher level tools for interpreting, checking and editing files.

%\todo{ES: don't focus so much on the CCL syntax -- this won't help
%  much in the long term.  restructuring the CCL itself will provide
%  more benefit, and should be explained in a paragraph.  (but isn't
%  the preceding text just this, explaining how higher level
%  abstractions could be introduced into the CCL?)}
%
%\todo{ES: could also re-investigate some of the current underlying
%  concepts of the CCL, e.g.\ being able to define a thorn interface
%  implicitly, without defining an explicit ``abstract interface''.
%  this reduces our inheritance model to comparing different thorns,
%  and no single location is entirely responsible for a given standard.
%  this has, in the past, often led to incompatibilities, when several
%  thorns agree on a standard, and then only some of them are changed,
%  while another has been ``left behind''.  on the other hand, not have
%  such standard-setting thorns keeps every thorn at the same level of
%  importance, without giving one thorn more importance that others
%  have to follow.  on the gripping hand, we're doing just this with
%  thorns such as ADMBase, so maybe we would actually want that.}
%
%\todo{ES: one important implementation issue is that we want to give
%  as many tools as possible access to the CCL database, both for
%  single thorns, for configurations, and for the (maybe inconsistent)
%  set of all thorns in a source tree.  we currently don't do this, and
%  using a standardised syntax may help here (think Eclipse,
%  GetComponents, etc.).}

A final consideration is the syntax for the CCL. Changing the CCL syntax could improve the 
ease with which the files could be constructed and edited, and importantly provide 
more options for standard tools which could be used to construct, investigate, debug and edit 
the CCL files. As an example, using a standardized syntax for CCL would allow users to take
advantage of the extensive features of the Eclipse Platform~\cite{eclipseweb}. Eclipse is an 
advanced Integrated Development Environment (IDE)\footnote{\url{http://en.wikipedia.org/wiki/Integrated_development_environment}}
 that includes features such as customizable syntax highlighting, auto-completion of code,
and dynamic syntax checking for languages it recognizes.
One option for revising the CCL syntax would be to use an existing data markup language that
incorporates metadata such as the Resource Description Framework (RDF)~\cite{rdfweb}. RDF is a 
widely used standard for describing data in internet tools. It uses URIs to describe the relationship
between two objects as well as the two ends of the link, which is commonly known as a \emph{triple}.
This would be a natural method for describing the dependencies between thorns, however RDF is
generally used as an extension of XML, which is not easily readable by humans. As the CCL files
must be generated by hand, it would be preferable to use an alternate format that focuses on
readability. One such example is YAML (YAML Ain't Markup Language)~\cite{yamlweb}, a data
serialization language with a strong emphasis on human readability. YAML represents data as a
series of sequences and mappings, both of which can be nested within others. While YAML does 
not inherently support metadata, it would be quite simple to add metadata to the thorns by adding
 extra mappings to the CCL files.

\section{Conclusion}

We have presented an overview of the Cactus Configuration Language (CCL) that describes Cactus thorns 
and have shown how the CCL is used in three different applications. 
The dependency information included in the CCL specification can be used to identify potential issues in designing 
complex codebases, and to build high--level tools to better assist users in constructing codes for particular applications. 

New features needed in the CCL specification have been identified, including support for more numerical methods, multiple physical 
models, user-defined structures, scientific metadata and to address the growing complexity of interfaces.


\section*{Acknowledgment}

The development of Cactus and the CCL has been a long term and ongoing effort with many contributors and funders. In particular 
we acknowledge the contributions of Gerd Lanfermann, Joan Mass\'{o}, Thomas Radke, and John Shalf, and funding from the National Science Foundation, 
Max-Planck-Gesellschaft, and Louisiana State University. We also acknowledge colleagues in the Einstein Toolkit Consortium 
whose thorns provide the motivation and core use case for this work. 
%\todo{ES: Usually one thanks people first, and acknowledges funding later.} 

Work on thorn dependencies  was funded by NSF
\#0904015 (CIGR)
%\todo{ES: XiRel?}
and NSF \#0721915 (Alpaca). 
Eric Seidel acknowledges support from 
the NSF REU program (\#1005165) and thanks the Center for Computation \& Technology for hosting his undergraduate internship. 

%\todo{ES: Please acknowledge LONI and TeraGrid allocations that you
%  used.  For SimFactory, we say ``We accessed HPC resources on the
%  TeraGrid via allocation TG-MCA02N014, at NERSC supported by DOE
 % contract DE-AC02-05CH11231, and on LONI under the allocations
 % loni\_cactus and loni\_numrel.''}

\bibliographystyle{amsplain-url}

\bibliography{CCL,publications-schnetter,references}

\providecommand{\bysame}{\leavevmode\hbox to3em{\hrulefill}\thinspace}
\providecommand{\MR}{\relax\ifhmode\unskip\space\fi MR }
\providecommand{\MRhref}[2]{%
  \href{http://www.ams.org/mathscinet-getitem?mr=#1}{#2}
}
\providecommand{\href}[2]{#2}
\begin{thebibliography}{10}

\bibitem{ES-carpetweb}
Mesh Refinement with {Carpet}, URL \url{http://www.carpetcode.org/}.

\bibitem{chomp}
Pawel~Pilarczyk et. al., \emph{Chomp}, http://chomp.rutgers.edu.

\bibitem{Goodale02a}
T.~Goodale, G.~Allen, G.~Lanfermann, J.~Mass{\'o}, T.~Radke, E.~Seidel, and
  J.~Shalf, \emph{The {C}actus framework and toolkit: Design and applications},
  Vector and Parallel Processing -- VECPAR'2002, 5th International Conference,
  Lecture Notes in Computer Science (Berlin), Springer, 2003.

\bibitem{numhomology}
Seth Major, David Rideout, and Sumati Surya, \emph{Stable homology as an
  indicator manifoldlikeness in causal set theory}, Class.Quant.Grav.
  \textbf{26} (2009), no.~175008, eprint {0902.0434 [gr-qc]}.

\bibitem{eclipseweb}
{Eclipse}:~An open~development platform, URL \url{http://www.eclipse.org/}.

\bibitem{1456196}
Erik Schnetter, \emph{Multi-physics coupling of einstein and hydrodynamics
  evolution: a case study of the einstein toolkit}, CBHPC '08: Proceedings of
  the 2008 compFrame/HPC-GECO workshop on Component based high performance (New
  York, NY, USA), ACM, 2008, pp.~1--9.

\bibitem{Schnetter06a}
Erik Schnetter, Peter Diener, Nils Dorband, and Manuel Tiglio, \emph{A
  multi-block infrastructure for three-dimensional time-dependent numerical
  relativity}, Class. Quantum Grav. \textbf{23} (2006), S553--S578, eprint
  {gr-qc/0602104}, URL \url{http://stacks.iop.org/CQG/23/S553}.

\bibitem{Schnetter-etal-03b}
Erik Schnetter, Scott~H. Hawley, and Ian Hawke, \emph{Evolutions in {3D}
  numerical relativity using fixed mesh refinement}, Class. Quantum Grav.
  \textbf{21} (2004), no.~6, 1465--1488, eprint {gr-qc/0310042}.

\bibitem{TG_CRL}
Eric~L. Seidel, Gabrielle Allen, Steven Brandt, Frank L\"{o}ffler, and Erik
  Schnetter, \emph{Simplifying complex software assembly: the component
  retrieval language and implementation}, TG '10: Proceedings of the 2010
  TeraGrid Conference (New York, NY, USA), ACM, 2010, pp.~1--8.

\bibitem{ES-simfactoryweb}
{SimFactory}: Herding Numerical Simulations, URL
  \url{http://www.cct.lsu.edu/~eschnett/SimFactory/}.

\bibitem{yamlweb}
The Official {Y}AML~Web Site, URL \url{http://yaml.org}.

\bibitem{graphvizweb}
{Graphviz} Graph~Visualization Software, URL \url{http://www.graphviz.org}.

\bibitem{causets}
Rafael~D. Sorkin, \emph{Causal sets: {D}iscrete gravity}, Lectures on Quantum
  Gravity, Proceedings of the Valdivia Summer School, Valdivia, Chile, January
  2002 (A.~Gomberoff and D.~Marolf, eds.), Plenum, 2005, eprint
  {gr-qc/0309009}.

\bibitem{rdfweb}
{R}DF Semantic~Web Standards, URL \url{http://www.w3.org/RDF}.

\bibitem{CBHPC_SimFactory}
Michael Thomas and Erik Schnetter, \emph{Simulation factory: Taming application
  configuration and workflow on high-end resources}, CBHPC '10 (New York, NY,
  USA), ACM, accepted.

\bibitem{Cactusweb}
{Cactus}~{Computational} {Toolkit}, URL \url{http://www.cactuscode.org}.

\bibitem{einsteintoolkitweb}
The~{Einstein} {Toolkit}, URL \url{http://www.einsteintoolkit.org}.

\end{thebibliography}

\end{document}